# The viscosity-radius relationship for concentrated polymer solutions

Dave E. Dunstan

*Department of Chemical Engineering, University of Melbourne, VIC 3010, Australia.*

davided@unimelb.edu.au

A key assumption of polymer physics is that the random chains polymers extend in flow. Recent experimental evidence has shown that polymer chains compress in Couette flow in a manner counter to expectation. Here, scaling arguments developed previously are used to determine the relationship between the viscosity and chain radius of gyration, $R_G$. Scaling arguments determine the viscosity-radius of gyration relationship to be $\eta \sim R_G^9$. The viscosity is shown to be a power law function of the radius, and to decrease with decreasing radius under conditions where the chains are ideal random walks in concentrated solution. Furthermore, this relationship is consistent with both the widely observed viscosity-temperature and viscosity-shear rate behavior observed in polymer rheology. The assumption of extension is not consistent with these observations as it would require that the chains increase in size with increasing temperature. Shear thinning is thus a result of a decreasing radius with increasing shear rate as $R_G \sim \dot{\gamma}^{-n/9}$ where $\dot{\gamma}$ is the shear rate and n the power law exponent. The thermal expansion coefficient determines the variation in the power law exponents that are measured for different polymer systems. Typical values on n enable the measured reduction in coils size behavior to be fitted. Furthermore, the absurd notion that polymer chains extend to reduce the viscosity implies that an increasing chain size results in a reduced viscosity is addressed. This assumption would require that the viscosity increases with reducing coil radius which is simply unphysical.

**Keywords:** Polymer dynamics, viscosity-radius, concentrated polymer solutions

## 1. INTRODUCTION

Polymers are of interest for reasons spanning wide practical application, through to the elegant theories of polymer dynamics introduced by Kuhn, Flory and de Gennes.[1-3] The theories show universal applicability through scaling arguments, suggesting a reductionist truth that is the aspiration of many other branches of physics.[3] A key focus of the area has been to understand the molecular basis of rubber elasticity and polymer rheology.[4-6] Considerable effort has been devoted to developing models to predict the visco-elastic flow behavior of polymer solutions and melts.[7,8] A general behavior of the type shown in Figure 1 is observed for these systems where the viscosity is seen to decrease with increasing shear rate in classical visco-elastic behavior.[4,6,9,10] Prediction of this behavior has been undertaken using a number of molecularly based models with varying degrees of success. A simple empirical model, the Power Law model, has been used to model the shear thinning behavior with a power law exponent.[4,9,11] Typical power law exponents have been observed for polymer solutions that are in the range of 0.5 – 1.0 suggesting a key universal physics underlies this behavior.[4,9,12]

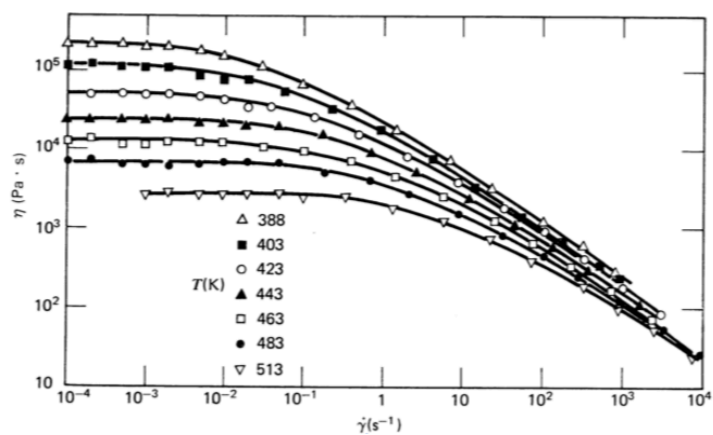

Figure 1. The viscosity versus shear rate behavior for a range of temperature for low density polyethylene melts at a range of temperatures. Original source, J. Meissner, *Kunststoffe*, 61, 576-582, 1971. Reproduced from *Dynamics of Polymeric Liquids I: Fluid Mechanics* by Bird et al.[6] with permission John Wiley and Sons.

A key assumption used in the models of polymer dynamics and rheology is that the chains extend in the flow to reduce the viscosity and imbue the solution with elasticity.[10] Here we present a brief history of the developments in polymer experiments and modelling in order to explain why the assumption that the chains extend in flow has become universally accepted and that there is little evidence for extension other than that observed for DNA and indeed recent experimental evidence showing contraction in flow.[13-25]

Here we review the chronology of the developments in experimental and theoretical developments in polymer physics. Gough (1805) and later Joule (1859) were the first to report on the contraction of rubber materials with increasing temperature.[26,27] Considerably later in 1920, Staudinger reported on the connected, polymeric, nature of rubber like materials that leads to their rather fascinating properties.[28,29] Kuhn was the first to postulate the dumbbell model of polymers in flow in his seminal 1933 Kolloid Z. paper.[1] He was the first to recognize that the macroscopic properties could be predicted from understanding single chain behaviour.[1] The dumbbell model first posed by Kuhn, in which the polymers behave as two beads on an elastic (Hookean) spring, is still used currently in modified forms.[5] The beads experience a Stokes drag that causes extension and compression as the dumbbell precesses in Jeffrey orbits.[30] In a later paper, Kuhn developed a statistical mechanical model to predict the Hookean force law for the chains that acts as a restoring force to counter the hydrodynamic forces.[31] Kuhn also predicted that the dumbbell could either extend or compress in flow as it rotates around the vorticity axis.

Several papers by Mooney[32,33], James and Guth[34] and Flory[35,36] then developed the Rubber Theory from statistical mechanics. The rubber theory accounts for both compression and extension[37].

In 1942 Kuhn and Grun published the first paper to assume that the chains extend in simple flow. They calculated the relationship between the reduced shear rate and the end-to-end vector of the chains.[38] In this paper the predicted reduced extension is plotted versus the reduced shear rate to show a limiting extension at high shear rates.

The 1950's then saw two key papers published by Rouse and Zimm.[39,40] The paper by Rouse presents the so called free draining model for the polymer coils where both hydrodynamic interactions and excluded volume are neglected. Zimm included hydrodynamic interactions for the theta condition (no excluded volume) in his 1956 paper.[40] Following the work of Rouse and Zimm, Peterlin published two papers that modelled the chains as an "ellipsoid whose axial ratio increases with shear".[41-44]

Flory published two seminal texts in 1953 & 1969 that examined polymers from a largely theoretical perspective and established the idea of excluded volume effects and "ideal chains" at high concentrations.[2,45]

The French physicists then followed by publishing a number of key papers that culminated in De Gennes' book *Scaling Concepts in Polymer Physics* in 1978 that introduced the reptation concept and further developed the scaling arguments first postulated by Flory.[3] Shortly after de Gennes book, Doi and Edwards authored *The Theory of Polymer Dynamics* in 1986.[46] The connection to the rheological behaviour was further advanced by Ferry in 1980[11] and then Bird et al. in their classic two volume text of 1987.[5,6] An elegant and comprehensive review by Larson[10] from 2005 outlines the development of the field in a chronological manner.

The experimental developments in the field, as reviewed below, have generally been coincident with advances in experimental methods that have enabled measurement of the key the parameters. Two of these are the chain deformation and orientation in flow. The advent of lasers and the development of light scattering with high resolution enabled the first light scattering measurements in flow to be undertaken. The pioneering work of Cottrell, Merrill and Smith in 1969 was the first measurement of light scattering from polymer solutions in shear[47]. More recently, Link and Springer 1993[48] and then Lee, Solomon[49] and Muller 1997 furthered this work.[50] Generally, the interpreted deformation is much less than the models predict. Much of the observed behaviour can be interpreted as orientation of the random ensemble of prolate chains in the flow field.[49,51,52] The orientation of the prolate chains, in Jeffrey orbits, increases the scattering cross section in the direction perpendicular to the vorticity direction resulting in the appearance of extension. The quiescent solution is isotropic and becomes anisotropic through flow induced orientation of the prolate chains. Rheo-optic measurements on dilute solutions of polydiacetylenes in Couette flow show increased projection of the chains in the flow direction, with no deformation of the backbone.[49]

A significant body of work has been generated measuring the flow induced birefringence of polymer solutions. This field is rather neatly summarised by Meissner and Janeschitz-Kriegl.[53] Much of the work has focused on measuring the stress optic coefficients and validating the so called stress-optic law. It should also be noted that the stress optic coefficients are found to be both positive and negative for differing polymer systems, suggesting that the flow induced behaviour is very different for the systems measured.[53] The variation in the stress-optic coefficients arises from the inherent refractive index difference between the backbone and the solvent, the flow induced orientation of the prolate chains and the spatial orientation of the chains.[46]

More recently a number of very elegant works on fluorescently labelled DNA in flow have been undertaken. Two of the key papers are by Smith, Babcock and Chu, Science 1999[16] and Le Duc, Haber, Bao and Wirtz, Nature 1999.[20] In both works the DNA was visualised using fluorescence microscopy with sliding plates to generate Couette flow and maintain the DNA molecules in the field of view. The problem with these works is that the DNA is claimed to be representative of random chain polymers in solution. The images show that this is not the case. The DNA images are not of a random chain whose conformation is determined by entropy.[16,20] Furthermore, the resolution of the microscopy method determines that compression is difficult to observe.[16] Larson has written a comprehensive review of the rheology of dilute solutions of flexible polymers focusing on the progress and problems.[10] A considerable component of the review is focused on simulations and modelling the data obtained from DNA. A key conclusion is that the measured deformation is less than expected. Given the importance of understanding polymer deformation in flow, the body of experimental data is not perhaps as comprehensive as would be expected with considerable weight being given to DNA. It should be noted that DNA does not show the same rheological behaviour as that observed for typical random coil polymers. Typical random coil polymers have conformation that is determined by their entropy, show decreasing viscosity with increasing shear rate and increasing temperature.[54, 55] Calf thymus DNA shows decreasing viscosity with shear rate with increasing viscosity with temperature.[55] A recent study by Bravo-Anaya et al. has shown that the rheological behaviour is a result of interacting aggregates of the DNA molecules in flow.[54] The interaction between the DNA molecules is suggested to be driven by H-bonding.

Rheo-optical measurements on synthetic polymers have shown chain orientation in dilute solution and compression at concentrations above critical overlap in the semi-dilute region.[23, 49, 56] The semi-dilute region is defined as where the chain interact with excluded volume effects present. The concentrated region is such that the chains behave as ideal random walks and their size scales as the square root of the molecular weight. Compression in flow has been observed for semi-dilute polymers in Couette flow. These experimental results have prompted a revision of the idea of extension being a universal assumption for polymers in simple planar flow. An alternative approach that assumes compression, allows the measured radius-shear rate behaviour to be predicted, and the power law behaviour observed for polymers in flow to be modelled.[25] Furthermore, using a force balance argument that predicts the shear thinning rheological behaviour, also enables the viscosity-radius relationship to be predicted. The predicted power law behaviour of the viscosity-radius is in close agreement with the experimentally observed behaviour.[25] Interestingly, this shows that the viscosity decreases as the radius decreases in a manner that is physically consistent with the observed behaviour for concentrated random chain polymers.[25]

Since Kuhn's original paper, the possibility of compression in Couette flow has not been considered and extension is assumed in the field.[3-6, 46, 57] The compression component has been ignored for chains in flow, however, recent experimental evidence has shown chain compression in Couette flow at semi-dilute concentrations.[22, 23, 25]

Here, scaling arguments are used to show that chains are predicted to compress in flow, and that variations in the observed power law exponents for shear thinning can be explained by the non-ideality embodied in the thermal compressibility of differing polymers.[58, 59]

## 2. THEORY

Adam and Delsanti first used scaling arguments to derive the viscosity-temperature relationship for semi-dilute polymer solutions [60]:

$$\eta \sim T^{-\frac{9(2-3\nu)}{3\nu-1}} \sim T^{-9} \qquad [1]$$

Here, the exponent of -9 is obtained using the scaling exponent $\nu = \frac{1}{2}$ as is found for the concentrated polymers in good solvents.[3, 45] Equation 1 is assumed valid for concentrated polymer solutions as the derivation by Adam and Delsanti is of a general nature and not restricted to the semi-dilute region providing C>C*. Furthermore, the viscosity of concentrated solutions decreases with increasing temperature (see figure 1) with a power law behavior that is consistent with the $T^{-9}$ predicted by Equation 1. Cheng et al. and Daoud et al. have shown that the scaling exponent varies from 3/5 to 1/2 in the semi-dilute regime.[58, 61, 62] Here the value of $\nu = \frac{1}{2}$ is used as the accepted value for chains in concentrated solution where excluded volume effects may be ignored. The justification for this assumption is that in concentrated solutions the chains are interacting and the excluded volume interactions become isotropic such that they can effectively be ignored.

Assuming that the chains are ideal, and that the entropic force determines the chain response to an external force, yields the usual Hookean force law:

$$f_s = 3k_B Tr/R_0^2 \qquad [2]$$

Where $f_s$ is the entropic force, $k_B$ Boltzmann's constant, T the absolute temperature, r the average end to end distance of the chain, and $R_0$ the end-to-end distance of the unperturbed chain. Here, it is noted that the mean square of end-to-end distance of the chain is related to the radius of gyration such that $<r^2> = 6R_G^2$ where $R_G$ is the radius of gyration.[57] Debye first derived the relationship between the end-to end vector and the radius of gyration.[63] Flory presents the complete argument for random chains of large molecular weight in *Statistical Mechanics of Chain Molecules* (p5 Eq 6.)[2] The end-to-end vector is the sum over all the segments. The radius of gyration is the root mean square distance of the collection of masses from their centre of gravity. Lagrange (appendix A in Flory)[2] related the centre of gravity for a system of masses to the distances between their centres taken pairwise. Providing the number of segments is large, the result presented in the paper is valid.

It is assumed herein that in the scaling relationships the end-to-end distance and the radius of gyration are equivalent.

Equation 2 shows that for a given force, an increase in temperature results in a decreasing end-to-end distance of the chain, and therefore radius. This is true for polymeric materials (rubbers) under tension, where contraction with increasing temperature is observed.[26,27,37,64-66] Remarkably, the contraction of rubber materials with temperature is predicted by the theory and has been observed experimentally. This effect was a key finding of the early work of Gough and Joule that was at the time not explained.[26,27]

Physical measurements on polymeric materials show that the chains contract with increasing temperature, as predicted by the entropic models of polymer chains developed from statistical mechanics.[2,3,31,67,68] The book by Mark and Erman, *Rubberlike Elasticity*, gives a comprehensive review of the measurement and interpretation of the mechanical measurements on rubbers.[68] Several key papers on the measurement and interpretation are by Shen et al. and Anthony et al.[69,70] The viscosity of concentrated polymer solutions is seen to decrease with increasing temperature suggesting that the chains contract with temperature.[6,11] Equation 2 can then be used to derive the temperature radius relationship that is in accord with the experimental data.[37,64] It is assumed that the form of Equation 2 is correct for a concentrated ensemble of chains. The form of the equation describing rubber elasticity relates the shear modulus of the material to the temperature:

$$G = NkT \qquad [3]$$

Where N is the density of entanglements. The definition of the modulus indicates that a concentrated polymer will contract as the temperature increases, as is experimentally observed.[37, 64, 68, 71] Maintaining the system under constant stress while varying the temperature requires that the entangled system will show an inverse strain relationship with temperature. Assuming that the entangled system will deform affinely, the strain will be proportional to the change in end-to-end distance.

At steady state the compressive and entropic forces on the chain will be equal and constant. Assuming that the force is approximately constant in Equation 2, then yields the relationship:

$$T \sim 1/r \qquad [4]$$

And as r ~ $R_G$ it follows that;

$$T \sim 1/R_G \qquad [5]$$

Combining [1] and [5] using the transitive property of equality yields:

$$\eta \sim R_G^9 \qquad [6]$$

Equation 6 shows that the viscosity decreases with decreasing radius of gyration when the entropic contribution to the chain restoring force is considered. Several approximations have been made in deriving this equation. Equation 5 is for a constant force and is valid at constant force. It is usually measured at an apparent equilibrium.[68] The viscosity is measured in steady state. The chains in flow will tumble in Jeffery orbits and therefore experience an average force as they tumble. It is assumed that they experience an averaged and effectively constant force in steady state flow. This power of the radius shown in Equation 6 suggests a cubic dependence on the volume fraction and thus a three-body interaction. Einstein proposed a volume fraction squared dependence of the viscosity at high concentrations.[57] The decreasing viscosity observed in typical shear thinning then results from a decreasing coil size in solution.[5, 11] Furthermore, by assuming that the polymeric solutions show power law behaviour [4, 9]:

$$\eta \sim \dot{\gamma}^{-n} \qquad [7]$$

Where $\dot{\gamma}$ is the shear rate. Experimentally measured values for n typically lie between 0.5 and 1. [9, 12]

Combining [6] and [7] yields:

$$R_G \sim \dot{\gamma}^{-n/9} \quad [8]$$

Equation 8 determines that as the shear rate increases, the radius decreases as the observed shear thinning occurs. This has been experimentally observed as is shown in Figure 2 below. The data of Figure 2 yields measured power law behavior for the decrease in radius with shear rate. For the two systems measured, the exponents are 0.07 (PMMA) and 0.0042 (BCMU) yielding power law exponents of n = 0.63 and 0.038 respectively. The power law exponent, n = 0.63 for PMMA is well within the range of values found for these polymer systems. Remarkably, the Power Law index for PMMA is found to be 0.62[72]

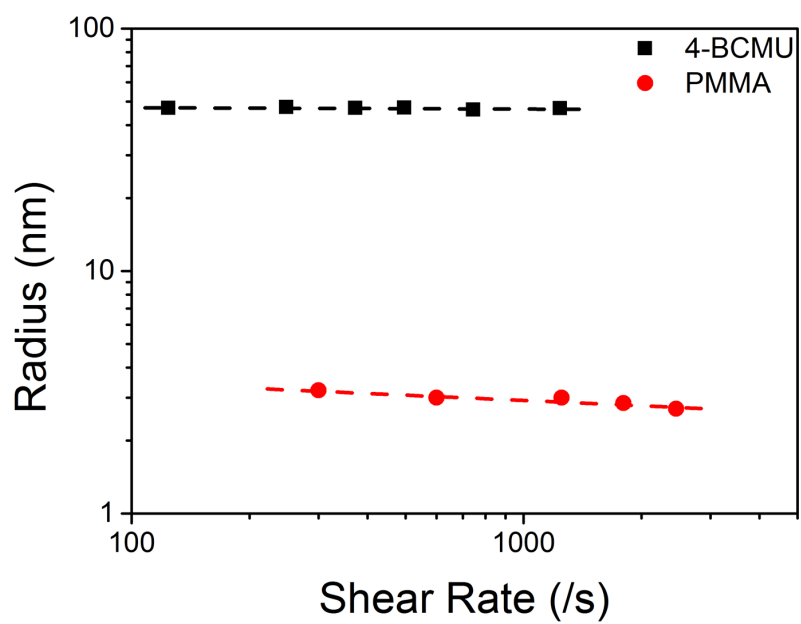

FIG 2. Measured end-to-end distance plotted as log r versus log shear rate. Data for 800kD 4-BCMU [23]. has the fitted equation: log r = -0.0046log($\dot{\gamma}$) + 1.7 with the coefficient of determination: $R^2$ = 0.23. Data for 49kD FRET tagged PMMA in Couette flow shows the fitted equation: log r = -0.072log($\dot{\gamma}$) + 0.69 with $R^2$ = 0.88. The lines of best fit yield an inverse 0.07+/-0.02 power of the radius with shear rate for the PMMA and 0.0042 +/-0.002 for the 4-BCMU. The error bars are approximately the size of the symbols. The error associated with each point is: ~5% in the shear rate due to the radius/gap ratio of the Couette cell. For the 4-BCMU the un-sheared size of the chain is 49+/-1nm and for PMMA the size is 4+/-0.1nm. Taken from Reference 25: D. Xie, D.E. Dunstan (2017) Modelling polymers as compressible elastic spheres in Couette ow. *Substantia* 1(1): 43-48. doi: 10.13128/Substantia-10, FIG 1.

A considerable body of work exists in the literature relating to the thermo-elasticity of polymeric materials.[37,58,65,67,68] Variations in the value of n, the power law exponent, is determined by the thermal expansion coefficient of the materials, that is attributed to the energetic interactions and therefore deviations from ideality and the universal physics described above. Mark, Price et al. and Flory present a thermodynamic argument relating the equilibrium force required to maintain a rubber strip at constant elongation as [2,45,65,68]:

$$f = \left(\frac{\partial U}{\partial L}\right)_{T,V} - T\left(\frac{\partial S}{\partial L}\right)_{T,V} \qquad [9]$$

Where $\partial U/\partial L$ $is$ the derivative of the internal energy of the chains with respect to length and $\partial S/\partial L$ is the derivative of the entropy. The two terms in equation [9] are then the energetic and entropic contributions to the total force. The total force is then simply written as:

$$f = f_e + f_s \qquad [10]$$

where $f_e$ and $f_s$ are the energetic and entropic contributions to the restoring force.

It can be shown for Gaussian chains that [65,73]:

$$\frac{f_e}{f} = T\frac{d\ln<R_0^2>}{dT} \qquad [11]$$

Where the thermal expansion coefficient κ is defined;

$$\kappa = d\ln r^2/dT \qquad [12]$$

A number of studies have reported experimental values of κ for a range of different polymer and solvent systems. The recent literature is summarized by Graessley and Fetters.[58,59] The κ values reported for the different polymer systems can be either positive or negative indicating that the enthalpic intramolecular thermal expansion deviates from Equation 5 both positively and negatively.[68]

Integrating [12] and ignoring the constant terms yields:

$$\ln R_G^2 \sim \kappa T \qquad [13]$$

Combining equations [1] and [13] yields:

$$\eta^{-9} \sim \ln R_G^{2/\kappa} \quad [14]$$

Equation [13] shows that the energy contribution to the change in chain size with viscosity depends on the sign of κ. The energy contribution adds to the entropic contribution so that:

$$\eta \sim R_G^9 + \left(\frac{1}{\frac{2}{\kappa}\ln R_G}\right)^9 \quad [15]$$

Where the sign of κ determines whether the second term is added or subtracted from the first term in equation [15].

Using the first term of the Taylor expansion yields:

$$\eta \sim R_G^9 + \left(\frac{\kappa}{2(R_G-1)}\right)^9 \quad [16]$$

Equation 16 shows that the sign of the energy contribution to the thermal expansion coefficient determines the rate at which the viscosity changes with radius.

This equation then demonstrates in physical terms why the values of n vary for different polymers. The values of κ vary for different polymers in both sign and magnitude, and this variation yields differing power law exponents (n) and behavior.[58, 59] There has been some discussion in the literature as to the meaning of κ, however it is assumed to be intramolecular in nature and a result of the conformational energies of the network chains. Essentially it is the ability of the chemical bonds in the chain to absorb energy. Thus, it is independent of the concentration or swelling of the network, the polymer molecular weight, or degree of cross linking.[67, 68] It is a measure of the degree of non-ideality of the chains and as shown above, has implications for the power law behavior of the viscosity.[9, 12]

The model presented here is consistent with experimental observations of chain contraction with increasing temperature and the viscosity of semi-dilute polymer solutions decreasing with increasing temperature and increasing (Couette) shear rate.[22, 23, 56] Furthermore, in extensional strain where polymer extension has been measured, the extensional viscosity increases.[6, 11] While the measured extensional viscosity of concentrated polymer solutions are observed to increase with increasing extensional strain rate, there is currently no consensus on the power law behavior. This is due to the apparent lack of reliability of the data for the instruments used, however, the general observation is that the extensional viscosity increases with increasing extensional strain rate.[74] However, the extensional viscosity measurements suggest that the chains are increasing in size in extensional flow. Extension of chains in extensional flow has been measured using neutron scattering.[75-77] Equations 7 and 8 suggest that the increasing extensional viscosity is due to increasing chain size and that this behavior should be described by power law behavior and that the power law index, n, should be negative in the case of extension.

The equations derived from scaling arguments describe the temperature dependence of the chain size and viscosity. The extension of DNA in flow may also be rationalized where the temperature dependence of the viscosity is the opposite of that observed for typical random chain polymers. As such, the behavior of DNA may not be representative of "polymers" as is generally claimed.[16, 78] Nonetheless, the increasing viscosity with temperature found for DNA is consistent with the model developed herein.

One of the reviewers of the first version of the manuscript suggested that the same argument as that used for the temperature dependence to derive Equation 6 could be developed using the concentration dependence of the radius and viscosity. In the semi-dilute region, an inverse relationship between the viscosity and radius is determined using this argument. This appears to contradict the argument developed in the current manuscript. The author currently has no explanation for this apparent contradiction. However, in the concentrated region, the coil size (radius) is independent of the concentration and therefore the arguments presented above are valid and consistent with the concentration dependence in this region.

## 3. CONCLUSIONS

Scaling arguments, developed previously, are used to determine the dependence of the viscosity, η, on the radius $R_G$ for polymer chains in concentrated solutions as; $\eta \sim R_G^9$. This relationship is developed for the conditions of ideal random chains in concentrated solutions. The relationship derived is consistent with recent experimental observations and is also physically consistent with the shear thinning and temperature dependence observed for typical polymer solutions and melts. Shear thinning is thus a result of a decreasing radius with increasing shear rate as $R_G \sim \dot{\gamma}^{-n/9}$ where n is the power law exponent. The thermal expansion coefficient determines the variation observed in the shear thinning power law exponents for different polymer systems. Recent experimental evidence on DNA extension in flow is also consistent with the model developed where the viscosity of the DNA solutions increases with temperature in a manner that is not typical of random chain polymers. This suggests that DNA is not a universal model of random chain polymers as purported in the literature.

**ACKNOWLEDGEMENT.** I would like to thank Maja Dunstan and Greg Martin for proofing the manuscript. I would like to thank the reviewers for their constructive comments that have been incorporated into the revised manuscript.

## 4. REFERENCES

1. W. Kuhn, *Kolloid Zeitschrift*, 1933, **62**, 269-285.
2. P. J. Flory, Statistical Mechanics of Chain Molecules, Hanser Publications: New York, 1988.
3. P. G. deGennes, Scaling Concepts in Polymer Physics, Cornell University Press: Ithaca, 1979.
4. B. B. Bird, W. E. Stewart and E. N. Lightfoot, Transport Phenomena, John Wiley, NY, 2002.
5. R. B. Bird, C. F. Curtiss, R. C. Armstrong and O. Hassager, Dynamics of Polymeric Liquids, Volume II, Kinetic Theory, Wiley-Interscience: New York, 1987.
6. R. B. Bird; R. C. Armstrong and O. Hassager, Dynamics of Polymeric Liquids, Volume I, Fluid Mechanics, Wiley-Interscience: New York, 1987.
7. P. S. Doyle, E. S. G. Shaqfeh and A. P. Gast, *Journal of Fluid Mechanics*, 1997, **334**, 251-291.
8. D. Petera and M. Muthukumar, *Journal of Chemical Physics*, 1999, **111**, 7614-7623.
9. H. A. Barnes, J. F. Hutton and K. Walters, An Introduction to Rheology, Elsevier, Amsterdam, 1989.
10. R. G. Larson, *Journal of Rheology*, 2005, **49**, 1-70.
11. J. D. Ferry, Viscoelastic Properties of Polymers, John Wiley, NY, 1980.
12. R. A. Stratton, *Journal of Colloid and Interface Science*, 1966, **22**, 517-530.
13. H. P. Babcock, D. E. Smith, J. S. Hur, E. S. G. Shaqfeh and S. Chu, *Physical Review Letters*, 2000, **85**, 2018-2021.
14. T. Perkins, D. Smith, R. Larson and S. Chu, *Science*, 1995, **268**, 83-87.
15. T. T. Perkins, D. E. Smith and S. Chu, *Science*, 1997, **276**, 2016-2021.
16. D. E. Smith, H. P. Babcock and S. Chu, *Science*, 1999, **283**, 1724-1727.
17. D. E. Smith and S. Chu, *Science*, 1998, **281**, 1335-1340.
18. R. E. Teixeira, H. P. Babcock, E. S. G. Shaqfeh and S. Chu, *Macromolecules*, 2004, **38**, 581-592.
19. R. E. Teixeira, A. K. Dambal, D. H. Richter, E. S. G. Shaqfeh and S. Chu, *Macromolecules*, 2007, **40**, 2461-2476.
20. P. LeDuc, C. Haber, G. Bao and D. Wirtz, *Nature*, 1999, **399**, 564-566.
21. N. Y. Chan, M. Chen and D. E. Dunstan, *European Physical Journal E*, 2009, **30**, 37-41.

22. N. Y. C. Chan, M. Chen, X.-T. Hao, T. A. Smith and D. E. Dunstan, *J. Phys. Chem. Lett.*, 2010, **1**, 1912-1916.
23. D. E. Dunstan, E. K. Hill and Y. Wei, *Macromolecules*, 2004, **37**, 1663-1665.
24. D. E. Dunstan and Y. Wei, *European Physical Journal Applied Physics*, 2007, **38**, 93-96.
25. D. Xie and D. E. Dunstan, *Substantia*, 2017, **1**, 43-47.
26. J. Gough, *Memoirs of the Literary and Philosophical Society Manchester*, 1805, **1**, 288-325.
27. J. P. Joule, *Philosophical Transactions of the Royal Society of London*, 1859, **149**, 91-131.
28. H. Staudinger, *Ber. Dtsch. Chem. Ges.*, 1920, **53**, 1074-1083.
29. H. Staudinger, *Naturwiss*, 1934, **22**, 65-72.
30. G. B. Jeffrey, *Proceedings of the Royal Society of London Series A*, 1922, **102**, 161-179.
31. W. Kuhn, *Kolloid Z*, 1934, **68**, 2-16.
32. M. Mooney, *Journal of Applied Physics*, 1940, **11**, 582-592.
33. M. Mooney, *Journal of Applied Physics*, 1948, **19**, 434-444.
34. H. M. James and E. Guth, *Journal of Chemical Physics*, 1943, **11**, 455-481.
35. P. J. Flory and J. R. Jr., *The Journal of Chemical Physics*, 1943, **11**, 521-526.
36. P. J. Flory and J. R. Jr., *The Journal of Chemical Physics*, 1943, **11**, 512-520.
37. L. R. G. Treloar, The Physics of Rubber Elasticity, Oxford University Press, Oxford 1975.
38. W. Kuhn and F. Grün, *Kolloid-Zeitschrift*, 1942, **101**, 248-271.
39. P. E. Rouse, *J. Chem. Phys.*, 1953, **21**, 1272-1280.
40. B. H. Zimm, *J. Chem. Phys.*, 1956, **24**, 269-278.
41. A. Peterlin, *Journal of Polymer Science*, 1954, **12**, 45-51.
42. A. Peterlin, *The Journal of Chemical Physics*, 1963, **39**, 224-229.
43. A. Peterlin, *Kolloid-Zeitschrift und Zeitschrift für Polymere*, 1963, **187**, 58-59.
44. A. Peterlin, W. Heller and M. Nakagaki, *The Journal of Chemical Physics*, 1958, **28**, 470-476.
45. P. J. Flory, Principles of Polymer Chemistry, Cornell University Press, Ithaca, 1953.
46. M. Doi and S. F. Edwards, The Theory of Polymer Dynamics, Clarendon Press: Oxford, 1986.
47. F. R. Cottrell, E. W. Merrill and K. A. Smith, *Journal of Polymer Science Part A-2: Polymer Physics*, 1969, **7**, 1415-1434.
48. A. Link and J. Springer, *Macromolecules*, 1993, **26**, 464-471.
49. S. Gason, D. E. Dunstan, T. A. Smith, D. Y. C. Chan, L. R. White and D. V. Boger, *J. Phys. Chem. B.*, 1997, **101**, 7732-7735.
50. E. C. Lee, M. J. Solomon and S. J. Muller, *Macromolecules*, 1997, **30**, 7313-7321.
51. W. Bruns, *Colloid Polym. Sci.*, 1976, **254**, 325-328.
52. K. Solc and W. H. Stockmayer, *J. Chem. Phys.*, 1971, **54**, 2756-2757.
53. J. Meissner and H. Janeschitz-Kriegl, Polymer Melt Rheology and Flow Birefringence, Springer Berlin Heidelberg, 2012.
54. L. M. Bravo-Anaya, E. R. Macías, J. H. Pérez-López, H. Galliard, D. C. D. Roux, G. Landazuri, F. Carvajal Ramos, M. Rinaudo, F. Pignon and J. F. A. Soltero, *Macromolecules*, 2017, **50**, 8245-8257.
55. L. M. Bravo-Anaya, M. Rinaudo and F. A. Soltero Martínez, *Polymers (20734360)*, 2016, **8**, 1-19.
56. N. Y. Chan, M. Chen and D. E. Dunstan, *The European physical journal. E, Soft matter*, 2009, **30**, 37-41.

57. M. Rubinstein and R. H. Colby, Polymer Physics, Oxford University Press, 2003.
58. W. W. Graessley, Polymeric Liquids & Networks: Structure and Properties, Garland Science, New York, 2004.
59. W. W. Graessley and L. J. Fetters, *Macromolecules*, 2001, **34**, 7147-7151.
60. M. Adam and M. Delsanti, *J. Physique*, 1982, **43**, 549-557.
61. G. Cheng, W. W. Graessley and Y. B. Melnichenko, *Phys. Rev. Lett.*, 2009, **102**, 157801-157805.
62. M. Daoud, J. P. Cotton, B. Farnoux, G. Jannink, G. Sarma, H. Benoit, R. Duplessix, C. Picot and P. G. de Gennes, *Macromolecules*, 1975, **8**, 804-818.
63. P. Debye, *The Journal of Chemical Physics*, 1946, **14**, 636-639.
64. J. E. Mark, *Journal of Chemical Education*, 1981, **58**, 898-903.
65. C. Price, *Proceedings of the Royal Society of London. Series A, Mathematical and Physical Sciences*, 1976, **351**, 331-350.
66. J. E. Mark, *The Journal of Physical Chemistry B*, 2003, **107**, 903-913.
67. J. E. Mark, *Journal of Polymer Science: Macromolecular Reviews*, 1976, **11**, 135-159.
68. J. E. Mark and B. Erman, Rubberlike Elasticity: A molecular Primer, Cambridge University Press, Cambridge UK, 2007.
69. M. C. Shen, D. A. McQuarrie and J. L. Jackson, *Journal of Applied Physics*, 1967, **38**, 791-798.
70. R. L. Anthony, R. H. Caston and E. Guth, *The Journal of Physical Chemistry*, 1942, **46**, 826-840.
71. X. Shen, C. Viney, E. R. Johnson, C. Wang and J. Q. Lu, *Nat Chem*, 2013, **5**, 1035-1041.
72. E. Freire, O. Bianchi, E. E. C. Monteiro, R. C. Reis Nunes and M. C. Forte, *Materials Science and Engineering: C*, 2009, **29**, 657-661.
73. G. Allen, M. J. Kirkham, J. Padget and C. Price, *Faraday Transactions*, 1976, **67**, 1278-1292.
74. T. Sridhar, M. Acharya, D. A. Nguyen and P. K. Bhattacharjee, *Macromolecules*, 2014, **47**, 379-386.
75. J. Bent, L. R. Hutchings, R. W. RIchards, T. Gough, R. Spares, P. D. Coates, I. Grillo, O. G. Harlen, D. J. Read, R. S. Graham, A. E. Likhtman, D. J. Groves, T. M. Nicholson and T. C. B. McLeish, *Science*, 2003, **301**, 1691-1695.
76. M. Heinrich, W. Pyckhout-Hintzen, J. Allgaier, D. Richter, E. Straube, T. C. B. McLeish, A. Wiedenmann, R. J. Blackwell and D. J. Read, *Macromolecules*, 2004, **37**, 5054-5064.
77. P. Lindner and R. C. Oberthür, *Colloid & Polymer Science*, 1988, **266**, 886-897.
78. C. M. Schroeder, R. E. Teixeira, E. S. G. Shaqfeh and S. Chu, *Physical Review Letters*, 2005, **95**, 018301.